\documentclass{PoS}

\def\beq{\begin{equation}}
\def\eeq{\end{equation}}
\def\beqa{\begin{eqnarray}}
\def\eeqa{\end{eqnarray}}

\def\etal{{\sl et al.\/}}
\def\etc{{\sl etc.\/}}
\def\jhep{{\sl J.\ H.\ E.\ P.\/}}
\def\pr{{\sl Phys.\ Rev.\/}}
\def\prl{{\sl Phys.\ Rev.\ Lett.\/}}
\def\ptps{{\sl Prog.\ Theor.\ Phys.\ Suppl.\/}}

\title{Finding the critical end point of QCD:\\ lattice and experiment}

\ShortTitle{Critical Point: lattice and experiment}

\author{\speaker{Sourendu Gupta}\\
        TIFR, Mumbai\\
        E-mail: \email{sgupta@theory.tifr.res.in}}

\abstract{The current status of the search for the critical end
  point on the lattice and in experiments is discussed. A naive
  extrapolation is given of lattice results on the location of
  the critical end point to the continuum limit with the correct
  pion mass. An interpretation of current results on fluctuations
  is provided. A new method of analysis of experimental data is
  discussed, which provides a check on the removal of backgrounds,
  comparisons with QCD, and signals an approach to the critical
  end point.}

\FullConference{5th International Workshop on Critical Point and Onset of Deconfinement - CPOD 2009,\\
		 June 08 - 12 2009\\
		 Brookhaven National Laboratory, Long Island, New York, USA}

\begin{document}

\section{Lattice}

When the light quark masses are finite, as they are in the real world, the
critical point of QCD lies at finite temperature and chemical potential. As
a result, lattice computations are beset with the sign problem. A method
to bypass this and estimate the position of the critical end point (CEP) was
given in \cite{texp} and used in \cite{nt4,nt6,bint4}. The idea is to make
a Taylor expansion of the pressure---
\beq
   P(T,\mu_B) = \sum_n \frac1{n!}\chi^{(n)}(T) \mu_B^n,
\label{pressure}\eeq
where the Taylor coefficients, called the susceptibilities, are evaluated
at $\mu_B=0$, where there is no sign problem. The baryon number susceptibility
is the second derivative of the pressure with respect to $\mu_B$ and has the
expansion
\beq
   \chi_B(T,\mu_B) = \sum_n \frac1{n!}\chi^{(n+2)}(T) \mu_B^n.
\label{texp}\eeq
This susceptibility diverges at the CEP. The divergence can be
diagnosed using the series coefficients by the usual means and the CEP 
can be located. This is a physical point if the value of $\mu_B$
at the end point is real.

We implemented this on the lattice \cite{nt4,nt6} using two flavours of
light dynamical quarks. The quark mass was tuned such that the pion mass
was 230 MeV. The lattice cutoff, $\Lambda=1/a$ was varied between 800
MeV and 1200 MeV to estimate the range of lattice cutoff effects. All
lattice computations are done at finite spatial volumes. In this case
the spatial box size was between 4 fm and 6 fm near $T_c$. As a result,
the boxes were large in units of the pion Compton wavelength as well
as the mean thermal wavelength. The temperature scale was set using
three different renormalization schemes; the scale uncertainty is about
1\%. The simulation algorithm was the R-algorithm.  The main algorithmic
parameter, the molecular dynamics time step was changed by one order of
magnitude without any change in the results.

Several issues remain to be addressed---
\begin{enumerate}
\item What effect does the unquenching of the strange quark have? A claim
 that 2+1 flavour QCD has no CEP \cite{dfpnt4} turns out to be a lattice
 artifact, cured by decreasing the lattice spacing \cite{dfpnt6}.
 The numerical impact on the quark number susceptibility of unquenching
 the light quarks was earlier seen to be small \cite{unq}. This could
 indicate that unquenching the strange quark should not have a large
 effect on the position of the CEP.
\item The state of the art is to use a pion mass of about 230 MeV. Decreasing
 this towards the physical value of 140 MeV should have a numerical impact on
 the prediction of the location of the CEP. A quantitative estimate can be
 made using results presented in \cite{ray}.
\item The global structure of the phase diagram may be more complicated.
 We have nothing to say about this. Current lattice computations address
 only the phase transition closest to $\mu_B=0$.
\item Finally, the series expansion is only carried out to a finite
 order (8-th order in our case).  The limiting behaviour as the order is
 increased is intimately connected to finite volume effects. We discuss this
 next.
\end{enumerate}

All lattice studies are necessarily performed at finite volume. The finite
size scaling theory which is used to extrapolate to infinite volume is
well established in the usual case where the simulation can be directly
performed at the critical point. The maximum of the susceptibility then
diverges as a (positive) power of the volume (a large effect), and the
position of this maximum is shifted from its infinite volume limit with a
(negative) power of the volume (a small effect).  In this case, the
simulation cannot be carried out at the critical point and these large
and small effects have to be obtained (along with the position of the
critical point) by analyzing the series coefficients.

The result is simple. On any finite volume, $V$, the radius of convergence
first seems to approach a finite limit, $\mu_B^*(V)$, up to a finite order,
$n_*(V)$. Beyond this the radius of convergence will seem to diverge,
since $\chi_B$ is finite at all finite $V$. The large effect is that
$n_*(V)$ becomes infinitely large as $V\to\infty$. The small effect is
that $\mu_B^*(V)$ changes by a small amount in the same limit. In the
present day simulations the large effect is clearly visible, whereas the
small effect is still hidden in the statistical uncertainties.

Starting from present day lattice simulations, extrapolation to infinite
volume, zero lattice spacing and the physical pion mass, all taken together
would predict the most probable range for the location of the CEP to be
$T^E=165$--175 MeV and $\mu_B^E=250$--400 MeV. Note that there are many
uncertainties and caveats in each of the extrapolations. An experimental
search over a somewhat broader range of these parameters is therefore
advisable.

\section{Experiment}

Away from a critical point the correlation length, $\xi$, of baryon
number fluctuations is finite. As a result, in any volume $V$ there are
$N=V/\xi^3$ independently fluctuating sub-volumes. In the thermodynamic
limit, as $N\to\infty$, the fluctuations at temperature $T$ are Gaussian---
\beq
   P(\Delta B) \propto \exp\left(-\frac{(\Delta B)^2}{2VT\chi_B}\right),
   \qquad{\rm where}\qquad \delta B=B-\langle B\rangle.
\label{gauss}\eeq
One way to test whether the critical point is reached is to look for
deviations from such Gaussian behaviour \cite{fluct}. That such Gaussian
behaviour should be observable is the content of \cite{ebye}.

The current RHIC runs produce fireballs which freeze out in a region of
the phase diagram which is not expected to contain the CEP. If lattice
computations are correct about this, then one should see a clear signal
of non-critical behaviour in present data. One way to analyze these is
to construct the first few cumulants of the observed distribution, $[B^n]$
for $n\le4$, and to extract from these
the mean $\langle B\rangle =[B]$,
the variance, $\sigma^2=[B^2]$,
the skew, ${\cal S} = [B^3]/\sigma^3$,
and the Kurtosis ${\cal K} = [B^4]/\sigma^4$.
At a normal point on the phase diagram one must have
\beq
   \langle B\rangle\propto V,\quad
   \sigma\propto \sqrt V,\quad
   {\cal S}\propto 1/\sqrt V,\quad{\rm and}\quad
   {\cal K}\propto 1/V.
\label{scale}\eeq
In heavy-ion experiments the volume is not observable. So one is forced to
use a proxy. The STAR experiment in a recent analysis \cite{star} used the
number of participants as such a proxy; using this they verified the above
power-law scalings.

A clinching point would be to compare the microscopic cumulants with the
QCD expectations:
\beq
   [B^2] = (T^3 V) \left(\frac{\chi^{(2)}}{T^2}\right),\quad
   [B^3] = (T^3 V) \left(\frac{\chi^{(3)}}T\right),\quad
   [B^4] = (T^3 V) \chi^{(4)}.
\label{cumulants}\eeq
This is not possible until a few more questions are
clarified. First, have all non-thermal sources of fluctuations (minijets,
decays, \etc) been successfully removed? Answering this question needs a
complete control over the systematics of the analysis. This has not yet
been demonstrated. Next, at what stage of the evolution of the fireball were
the fluctuations set up (what values of $V$ and $T$ should one use in 
eq.\ \ref{gauss})?  This requires control over the theory of coupled
hydrodynamics and diffusion \cite{diffu}.

\begin{figure}
\begin{center}
   \scalebox{0.5}{\includegraphics{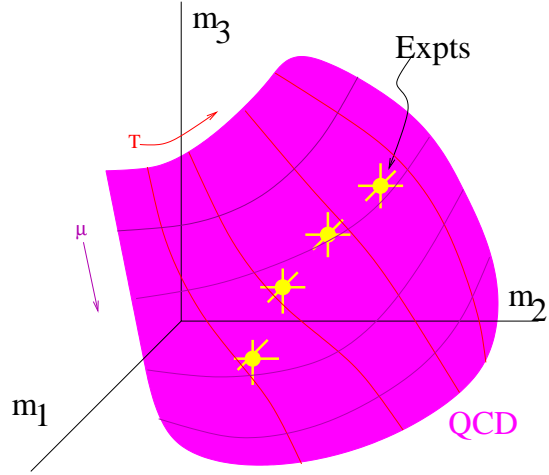}}
\end{center}
\caption{As $T$ and $\mu$ are varied, the QCD predictions lie on
  a surface in the space of measurements $(m_1,m_2,m_3)$. The
  data will lie on this surface if all non-thermal behaviour has
  been properly removed. In the happiest case, this could yield
  a comparison with QCD and a measurement of $T$ and $\mu$. Deviations
  from the surface {\sl only\/} in a small window of beam energy
  is a signal for the CEP.}
\label{fg.expt}
\end{figure}

In order to remove unmeasurable quantities like $V$ and $T$ from explicitly
appearing in the measurements one has to construct different combinations
of variables.  Once all backgrounds and systematics are under control,
the following combinations can be compared to QCD predictions---
\beqa
\nonumber
  m_1 &=& {\cal S} \sigma
     = \frac{\left[B^3\right]}{\left[B^2\right]} =
      \frac{\chi^{(3)}/T}{\chi^{(2)}/T^2} = \frac1{r_{23}},\\
\nonumber
  m_2 &=& {\cal K} \sigma^2
     = \frac{\left[B^4\right]}{\left[B^2\right]} =
      \frac{\chi^{(4)}}{\chi^{(2)}/T^2} = \frac2{r_{24}^2},\\
  m_3 &=& \frac{\cal K}{\cal S} \sigma
     = \frac{\left[B^4\right]}{\left[B^3\right]} =
      \frac{\chi^{(4)}}{\chi^{(3)}/T} = \frac2{r_{34}}.
\eeqa
These three measurements can be used to extract $\chi^{(2,3,4)}$ under
the assumption that all backgrounds have been removed and a comparison
with lattice QCD predictions \cite{nt6} is possible. However this
is a strong assumption.

A neater analysis with less bias is possible. Plot the experimentally
measured values of $(m_1,m_2,m_3)$ in a three-dimensional plot. Lattice
QCD predictions of these quantities can also be plotted in the same figure
(see Figure \ref{fg.expt}). As one varies $T$ and $\mu_B$, the theoretical
predictions trace out a surface. If the experimental points do not lie
on it, then non-thermal sources have not been completely removed. On
the other hand, if they do, then one can estimate the values of $T$
and $\mu_B$ the experiments correspond to by comparison with the QCD
predictions.

Near a critical point, one can never satisfy $V/\xi^3\gg1$. As a
result the central limit theorem will break down. In a static system
one would expect the Kurtosis to diverge. However the expansion
of the fireball rounds off the transition and renders $\xi$ finite
\cite{rajagopal}.  Nevertheless, $\xi$ has a maximum near the CEP. Since
${\cal K}\simeq\xi^3$ \cite{stephanov} the Kurtosis peaks near the
CEP. For the measurements suggested above, one would have
\beq
   m_1 \propto \xi^{(7-\eta)/2},\qquad
   m_2 \propto \xi^{5-\eta}, \qquad
   m_3 \propto \xi^{(5-\eta)/2}.
\label{fse}\eeq
In other words, all these quantities would have non-monotonic behaviour
as a function of the beam energy if the beam-energy scan passes through
the vicinity of the CEP. If desired, such analyses can be easily
extended to higher cumulants (when the system deviates from a Gaussian,
the higher cumulants become easier to measure).

One can use the plot of Figure \ref{fg.expt} in the search for the
CEP. Tune the background subtraction and cuts so that the present data
for $(m_1,m_2,m_3)$ lie on the QCD surface, as it should. Then in a
beam energy scan, the data will lie on the surface whenever the system
lies away from the CEP. In the vicinity of the CEP, however, effects
such as those discussed in \cite{rajagopal} drive the system away from
thermodynamic equilibrium. As a result, in this small window of energies
the experimental data will deviate from the surface, signaling critical
slowing down as a direct probe of the nearness of the CEP.


\begin{thebibliography}{99}
\bibitem{texp} 
  R.\ V.\ Gavai and S.\ Gupta, \pr, D 68 (2003) 034506.
\bibitem{nt4} 
  R.\ V.\ Gavai and S.\ Gupta, \pr, D 71 (2005) 114014.
\bibitem{nt6} 
  R.\ V.\ Gavai and S.\ Gupta, \pr, D 78 (2008) 114503.
\bibitem{bint4} 
  M.\ Cheng, \etal, \pr, D 79 (2009) 074505.
\bibitem{dfpnt4} 
  P.\ de Forcrand and O.\ Philipsen, \jhep, 0701 (2007) 077.
\bibitem{dfpnt6} 
  O.\ Philipsen, these proceedings.
\bibitem{unq} 
  R.\ V.\ Gavai and S.\ Gupta, \pr, D 73 (2006) 014004.
\bibitem{ray} 
  R.\ V.\ Gavai, S.\ Gupta and R.\ Ray, \ptps, 153 (2004) 270.
\bibitem{fluct} 
  M.\ A.\ Stephanov, K.\ Rajagopal, E.\ V.\ Shuryak, \pr, D 60 (1999) 114028.
\bibitem{ebye}
  M.\ Asakawa, U.\ W.\ Heinz and B.\ Muller, \prl, 85 (2000) 2072;\\
  S.\ Jeon and V.\ Koch, \prl, 85 (2000) 2076.
\bibitem{star} 
  B.\ Mohanty (STAR Collaboration), arXiv:0907.4476 [nucl-ex].
\bibitem{diffu} 
  D.\ T.\ Son and M.\ A.\ Stephanov, \pr, D 70 (2004) 0506001;\\
  S.\ Gavin and M.\ Abdel-Aziz, \pr, C 70 (2004) 034905;\\
  R.\ B.\ Bhalerao and S.\ Gupta, \pr C 79 (2009) 064901.
\bibitem{rajagopal}
  B.\ Berdnikov and K.\ Rajagopal, \pr, D 61 (2000) 105017.
\bibitem{stephanov}
  M.\ A.\ Stephanov, \prl, 102 (2009) 032301.
\end{thebibliography}
\end{document}